\documentclass[pre,aps,preprint]{revtex4-1}
\usepackage{graphicx}
\usepackage{amsfonts}
\usepackage{amssymb}
\usepackage{amsmath}
\usepackage{textcomp}

\usepackage{color}
\usepackage{psfrag}

\usepackage{epsfig}

\begin{document}

\title{Improved reconstructions of 
random media using dilation and erosion processes}

\author{Chase E.~Zachary$^{1}$ and Salvatore Torquato$^{1,2,3,4,5}$}
\email{torquato@princeton.edu}
\affiliation{$^{1}$Department of Chemistry, $^{2}$Department of Physics, 
$^{3}$Princeton Center for Theoretical Science, 
$^{4}$Program in Applied and Computational Mathematics,
and $^{5}$Princeton Institute for the Science and Technology of Materials,
Princeton University, Princeton, New Jersey 08544, USA}

\begin{abstract}
By using the most sensitive two-point correlation functions 
introduced to date, we reconstruct the microstructures of two-phase random media with heretofore
unattained accuracy. Such  media
arise in a host of contexts, including
porous and composite media, ecological structures, biological
media, and astrophysical structures. The aforementioned correlation functions are special
cases of the so-called {\it canonical} $n$-point correlation function $H_n$ 
and generalize the ones that have been recently employed to advance our ability
to reconstruct complex microstructures [Y. Jiao, F. H. Stillinger, and S. Torquato, Proc. Nat. Acad. Sci. {\bf 106}, 17634 (2009)].
The use of these generalized correlation functions is tantamount to dilating or eroding a reference phase of the target medium
and incorporating the additional topological information of the  modified media, thereby providing more accurate reconstructions of
percolating, filamentary, and other topologically complex microstructures.   We apply our methods
to a multiply-connected ``donut'' medium and a dilute distribution of ``cracks'' (a set of essentially zero measure) demonstrating improved
accuracy in both cases with implications for higher-dimensional and biconnected two-phase systems.      
The high information content of the   generalized two-point correlation functions suggests that it would be profitable
to explore their use to characterize the structural and physical properties of not only random media,
but also molecular systems, including structural glasses.

\end{abstract}

\maketitle



\section{Introduction}

Two-phase random media, which are partitions of space into two distinct
phases with interfaces that are known only probabilistically, abound in a host
of situations, including composite materials \cite{To02,Sa03,Zo06,Me07,Ga98}, cosmological \cite{Pe93,Le11} and geophysical
structures \cite{To02}, and biological systems \cite{Gi97,Ge08}. Importantly, the effective physical
and chemical properties of random media  are rigorously related to microstructural
correlations functions \cite{To02,Sa03}.   In principle, complete
microstructural information requires an enormously large set of such correlation functions,
including statistical descriptors describing higher-order correlations (e.g., three-point, four-point,
and higher-order statistics). Since in any practical setting only lower-order
correlation functions are available, it is of paramount interest to quantify the amount of
microstructural information contained in such reduced statistics and then to identify the most sensitive
descriptors.

Microstructure reconstruction techniques provide a means of probing these issues.
Reconstruction of a random medium that matches
limited structural information, i.e., finite set of targeted lower-order
correlation functions of the system, is an intriguing inverse problem \cite{Be87,Ye98}.
By comparing unconstrained correlation functions between the original and reconstructed medium,
one can quantify the capacity of the reduced statistics to specify the important 
structural features of the medium \cite{Ji10}.   
An effective reconstruction procedure enables one to generate
accurate renditions of random media at will, and subsequent
analysis can be performed on the reconstruction to obtain
desired macroscopic properties nondestructively \cite{To02, Ku06}.

Although there has been significant progress on reconstruction algorithms \cite{To02,Ro02,Ji09, Fu10},
it remains a challenge to reconstruct accurately percolating microstructures, sets
of essentially zero measure (e.g., cracked media and filamentary structures), and multiply
connected microstructures. In this paper, we introduce a  procedure
that enables one to begin to reconstruct such microstructures with heretofore
unattained accuracy.

It is noteworthy that a panoply of different types of correlation functions
arise in rigorous theories of structure-property relations for
heterogeneous media, which includes the $n$-point probability function $S_n$, $n$-point surface correlation functions, pore-size
functions, lineal-path functions, nearest-neighbor functions, among others \cite{To02}. It was shown that all of these
statistical descriptors are special cases of the more general
{\it canonical} $n$-point correlation function $H_n$, which is a hybrid
correlation-density function and can be  represented analytically
for a wide class of microstructures \cite{To86,To02}. Indeed, $H_n$ characterizes a {\it modified}
random medium that involves uniformly
``dilating" or ``eroding" a reference phase in the normal direction to the two-phase interface
in the actual random medium \cite{To02}.
{\it Dilation} is achieved by placing a ``test" sphere of radius $\delta$ into the non-reference
phase and determining the space available to the sphere as if the
reference phase was impenetrable to the test sphere. The space unavailable
to the test sphere is considered to be the modified or dilated reference phase.
The {\it erosion} process simply  reverses this procedure such that the reference
phase is allowed to be penetrated by a distance $\delta$ normal
to the two-phase interface. The quantity $H_n$ is an even more general quantity since
it allows for multiple test spheres, each of which has a different size; see  Refs. \cite{To02} and \cite{To86} for details.  Thus, $H_n$ contains considerably more information than any 
of the types of correlations functions that arise in the theory of
heterogeneous media. We will use special cases of $H_2$ in our reconstruction method, as described below.

 In practice, the enormous information 
content of the full $n$-point canonical function $H_n$ is neither experimentally 
nor theoretically accessible, and one must instead rely on \emph{reduced} statistical 
information about the microstructure.  It is 
an open and fundamental problem to identify via reconstruction techniques lower-order statistical
descriptors
that can be both manageably measured and yet reflect a substantial portion of the complete microstructural information contained in  $H_n$. The most prominent example of reduced 
statistical information in the two-point probability function $S_2({\bf r})$, probability of finding
two points separated by the displacement vector $\bf r$ in one of the phases. This function  is accessible from
scattering experiments \cite{DeBu49} or tomography \cite{RiToYeKe96}. Such information alone has been
shown to be generally insufficient to render accurate reconstructions \cite{Ye98,Ji10}.

The most successful reconstruction procedures to date have directly incorporated 
\emph{connectedness} information about the target media.  In  particular, Jiao, Stillinger and Torquato \cite{Ji09}
have shown that incorporation of the two-point cluster function $C_2$,
experimentally obtainable from tomography or other three-dimensional imaging techniques \cite{To02}, 
results in superior reconstructions.  The two-point cluster function defines the probability 
of finding two points in both  the same phase and in the same connected cluster of the 
microstructure \cite{To88, To02}.  
The function $C_2({\bf r})$ is the connectedness contribution to the  two-point
probability function, namely, $S_2(\mathbf{r}) = C_2(\mathbf{r}) + E_2(\mathbf{r})$, 
where $E_2$ is the two-point blocking function, corresponding to the case where two 
points fall within different clusters of the same phase.  
It has been shown that $C_2$ contains substantially
more information in excess to $S_2$ than even the three-point
probability function $S_3$, thereby highlighting the inefficiency of
incorporating additional information via the usual higher-order
correlation functions \cite{Ji09}.

However, there remains a large class of microstructures that are difficult to characterize
accurately using only $S_2$ and $C_2$.  Examples of such systems include 
heterogeneous media with percolating and/or multi-connected phases \cite{Qu97}, 
systems with zero-measure
structures such as cracked materials and filamentary structures 
that arise in the large-scale structure of the Universe \cite{Gi96}, and other topologically 
complex media. In particular, $C_2$ works best when the phase of interest
is below its percolation threshold; otherwise, it is a long-ranged function
that does not reflect structural details.
Moreover, when the phases occupy effectively a space of Lebesgue measure zero, 
the probability of finding a point ``within'' the set identically vanishes; these elements
are therefore unable to contribute to either $S_2$ or $C_2$.  

In this paper, we present an improved reconstruction procedure that naturally generalizes 
the $S_2$-$C_2$ methodology presented in Ref. \cite{Ji09}.  
Fundamental to our 
work is the information contained in the \emph{void space} external to the phase of interest \cite{To10,Za11}.  
In particular, we incorporate special and new  cases of $H_2$, which   contain additional
topological information associated with modifications of the target microstructure  via ``dilation'' and ``erosion'' processes,
thereby allowing a more accurate reconstruction procedure.   Importantly,
our methods contain previously-reported results for $S_2$ and $C_2$ as a special case, thereby 
providing a general means to study the amount of information contained in lower-order correlation 
functions.

\section{Generalized Two-Point Correlation Functions: Special cases of $H_2$}

Previous studies on reconstructions of heterogeneous media \cite{Ji09} have utilized the 
\emph{pore-size probability function} $F(\delta)$, defined as the probability of 
finding a spherical cavity of radius $\delta$ contained in the void space external 
to the phase of interest.  The pore-size probability function contains short-range
connectedness information within the void space and can also be interpreted as the 
volume fraction of the pore space after uniformly dilating the reference phase by a 
linear scale $\delta$ \cite{To10}; note that erosion by a linear scale
$\delta$ corresponds to the continuous extension of $F$ to 
negative values of $\delta$.  As such, it is a \emph{sensitive} one-point descriptor 
of heterogeneous media implicitly containing limited higher-order information.  

Higher-order pore-size probability functions are simply special cases 
of the canonical $n$-point correlation function $H_n$.  In particular, the 
\emph{$n$-point pore-size probability function} $F_n(\mathbf{r}^n; \{\delta\}_n)$ 
is the probability of finding $n$ spherical cavities of radii $\delta_1, \ldots, \delta_n$, 
all within the ``void" space external to the reference phase; in the notation of Ref. \cite{To86},
$F_n(\mathbf{r}^n; \{\delta\}_n) \equiv H_n(\mathbf{r}^n; \O; \O)$. Of particular interest 
here is the two-point pore-size probability function $F_2(\mathbf{r}; \delta)$
for a statistically homogeneous  medium, where $\mathbf{r} = \mathbf{r}_2-\mathbf{r}_1$
and $\delta = \delta_1 = \delta_2$.  This function 
also contains short-range connectedness information about the {\it modified} void space and can 
be interpreted as the two-point probability function $S_2$ for the void space after uniformly
dilating the reference phase by a linear scale $\delta$.  As such, it can be 
decomposed as $F_2(\mathbf{r}; \delta) = K_2(\mathbf{r}; \delta) + M_2(\mathbf{r}; \delta)$, where
$K_2$ and $M_2$ are the two-point pore-size cluster and blocking functions, respectively.

Since $F_2(\mathbf{r}; \delta) \rightarrow S_2(\mathbf{r})$ (for the void space) as $\delta\rightarrow 0$,
it follows that $F_2$ and its cluster counterpart $K_2$ rigorously contain more information 
about the microstructure than the standard $S_2$ and $C_2$ functions alone.  Thus, we incorporate
these generalized $S_2$ and $C_2$ functions in reconstructions, which amounts to 
applying dilation or erosion processes to modify the phase connectedness of a target microstructure. 
We then perform a
reconstruction procedure on the modified medium, which is expected to yield
improved reconstructions of the {\it original} unmodified medium.

\section{Reconstruction Procedure and Applications}

Our reconstruction procedure is a modification of the Jiao-Stillinger-Torquato methodology \cite{Ji09}, which 
extends the Yeong-Torquato simulated annealing algorithm \cite{Ye98} to include clustering information.
In particular, we overlay the image with a set of $N^d$ pixels taking values $1$ (within the 
phase of interest) or $0$ (in the void space), where $d$ is the Euclidean dimension.  By measuring the distances between 
void and surface pixels, we perform an initial dilation or erosion process to modify the 
microstructure prior to reconstruction.  Reconstruction of the modified structure
then proceeds by stochastic optimization
of the following objective (energy) function \cite{Ye98,To02}:
\begin{equation}
E = \sum_{\alpha=1}^M \sum_{j=1}^S \left[f_j(\mathbf{r}_\alpha) - f^{\prime}_j(\mathbf{r}_{\alpha})\right]^2,
\end{equation}
where $M$ is the number of sampling points and $S$  is the number of constrained correlation
functions $f_j$ with target values $f^{\prime}_j$.  
In this work we consider radially-averaged functions [e.g., $S_2(r)$], which are 
more readily accessible from scattering experiments and more relevant to studies 
in the thermodynamic limit \cite{Ji10}.  However, our methods can be directly 
extended to include anisotropic correlation functions.
The Jiao-Stillinger-Torquato procedure tracks the 
occurrence of ``cluster'' and ``surface'' events to update efficiently the cluster and boundary 
lists at each step of the optimization procedure \cite{Ji09}.  A cooling schedule
is chosen such that the energy $E$ approaches its
ground-state value within a very small tolerance \cite{Ji08}.

\begin{figure}[!t]
\centering
$\begin{array}{c c c}
\underset{\text{\Large (a)}}{\includegraphics[width=0.3\textwidth]{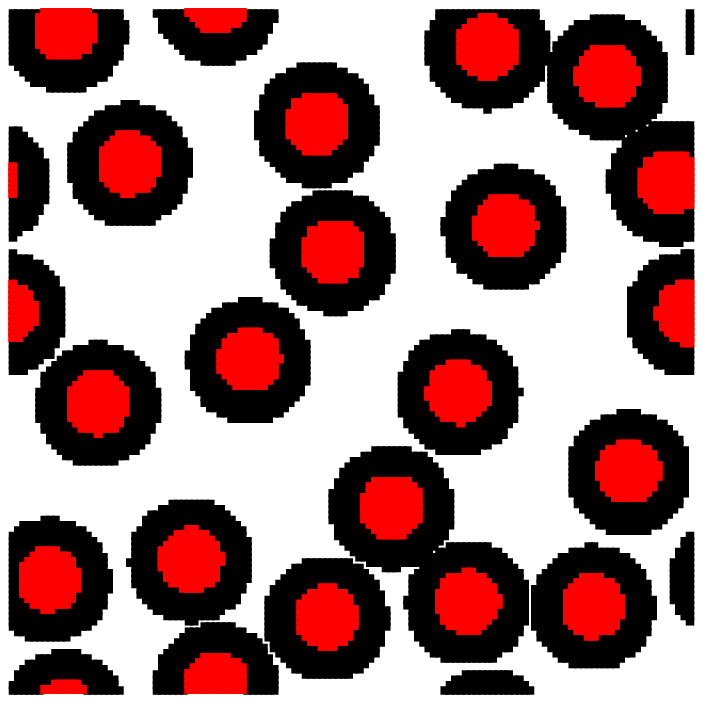}} &
\underset{\text{\Large (b)}}{\includegraphics[width=0.3\textwidth]{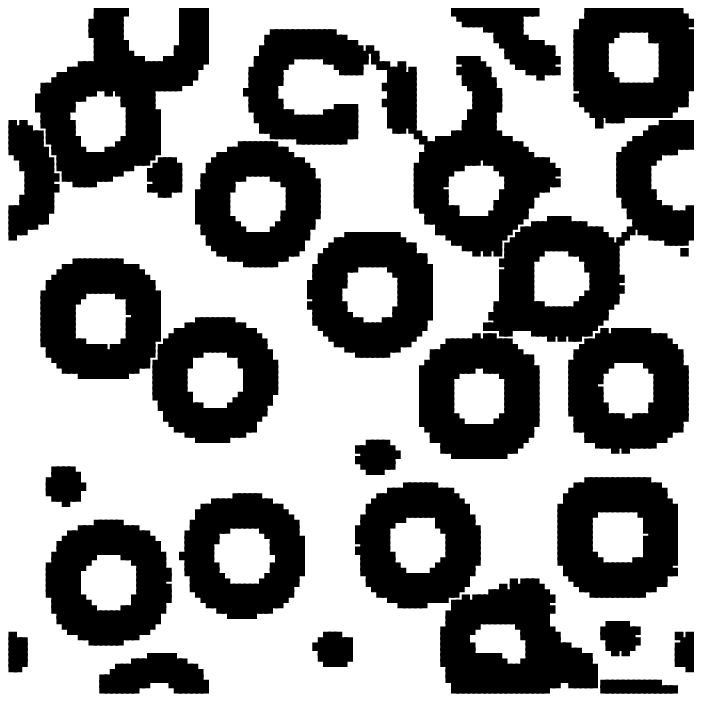}} &
\underset{\text{\Large (c)}}{\includegraphics[width=0.3\textwidth]{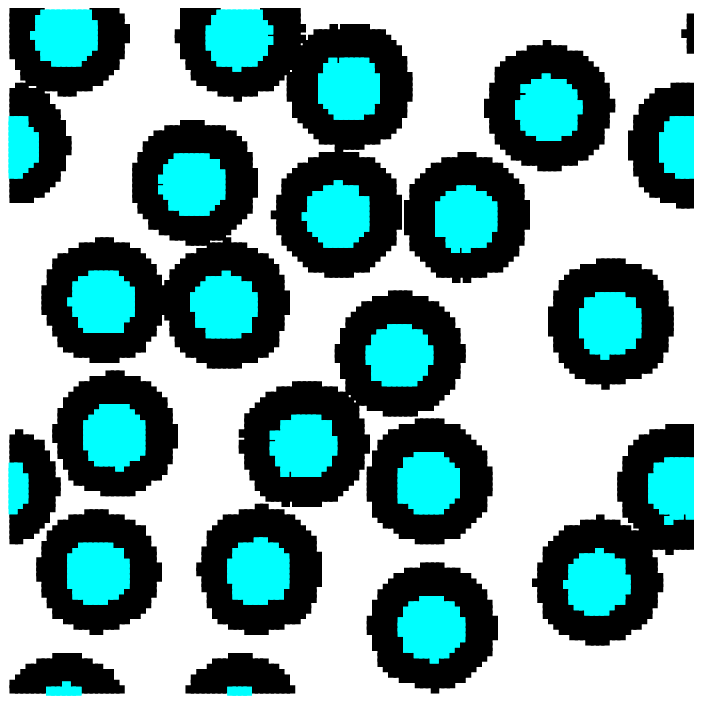}}
\end{array}$
\caption{(Color online)  The doubly-connected ``donut'' medium.
(a)  Original microstructure (outer rings) with modified target structure (inner disks; red online)
after an erosion process.  (b)  Direct $S_2(r)$-$C_2(r)$ reconstruction of the original microstructure.  
(c)  $S_2(r)$-$C_2(r)$ reconstruction of the modified target structure (inner disks; blue online) along with the 
final reconstructed system (outer rings) after a dilation process. All reconstructions
have a final energy (error) at most $E \sim 10^{-8}$.}\label{figone}
\end{figure}
We illustrate our improved reconstruction methodology by studying the
doubly-connected ``donut'' structure shown in Figure
\ref{figone}, consisting of annuli distributed in the plane.  It is noteworthy that nanoparticles
of mesoporous silica shells with hollow interiors have been designed with similar cross-sectional 
structures for applications to biomedical imaging and targeted drug delivery \cite{Ki08}.  The uniformity of the nanoparticle sizes and shapes,
including the hollow interiors, is particularly important and 
can be directly quantified using a combination of imaging techniques and pore-size distribution 
functions \cite{Ki08}.

The connectedness information of the donut structures is nontrivial because the two-dimensional
annulus is not a simply-connected topological object.  A direct reconstruction using $S_2$ and 
$C_2$ shows that, while the two-point cluster function is able to identify situations where
points fall within the same cluster, it cannot distinguish situations where the ``donut shells''
are closed and open.  Furthermore, isolated disks of particles appear in the reconstructed 
image with diameter approximately equal to the length over which $C_2$ is nonzero.  

\begin{figure}[!t]
\centering
$\begin{array}{c c}
\underset{\text{\Large (a)}}{\includegraphics[height=2.5in]{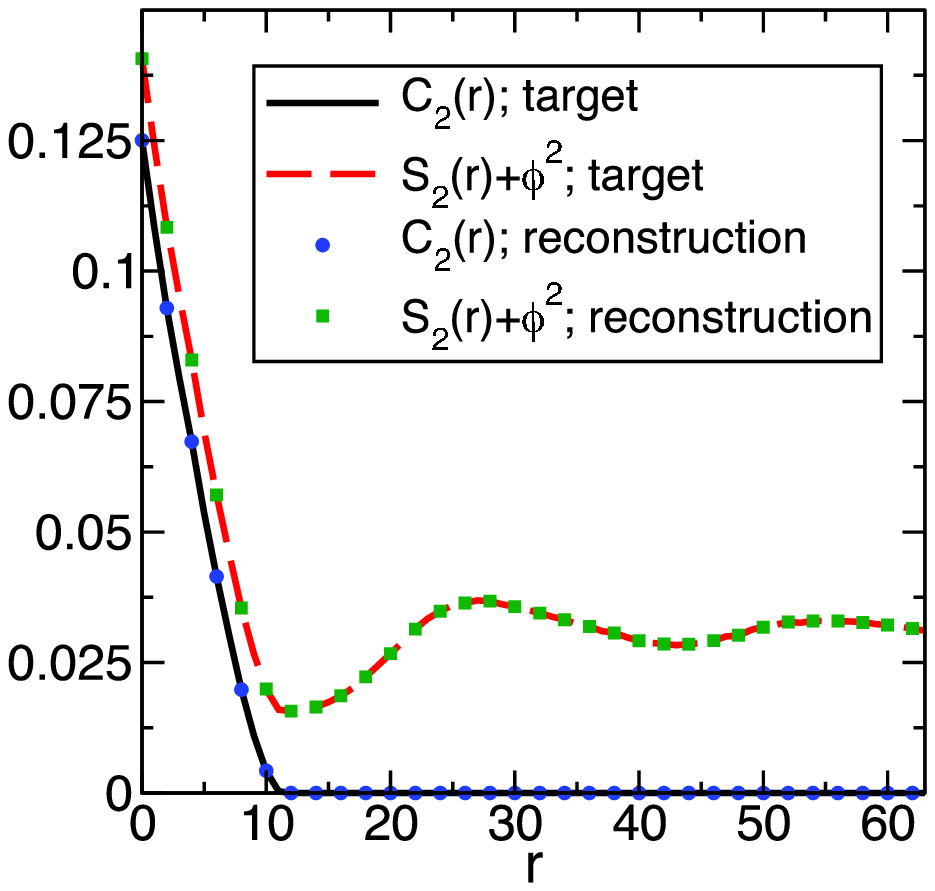}} &
\underset{\text{\Large (b)}}{\includegraphics[height=2.5in]{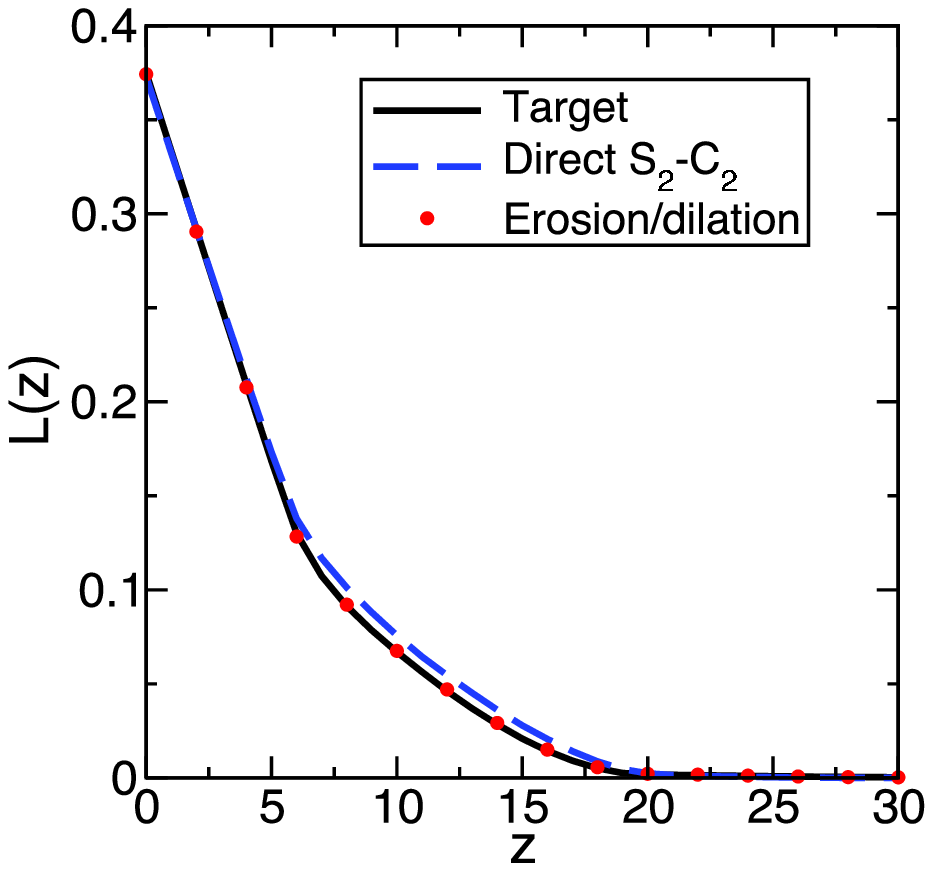}}
\end{array}$
\caption{(Color online)  
(a)  $S_2(r)$ and $C_2(r)$ for the modified target disk structure following an erosion process
and for the reconstructed disk microstructure.  (b)  Lineal path functions $L(z)$ for the original target donut 
structure, a direct $S_2(r)$-$C_2(r)$ reconstruction of the donut system, and the reconstructed
material using erosion and dilation processes.}\label{figonestats}
\end{figure}
We are able to improve the reconstruction procedure by targeting the spatial information contained
in the ``cores'' of the ring structures.  Although the annuli surrounding the cores are not 
simply-connected, by performing an erosion process on the initial microstructure, we generate
a ``modified'' system composed of simply-connected, spatially separated disks in the plane (see 
Figure \ref{figone}).  Reconstruction of the modified medium using $S_2$ and $C_2$ is able to reproduce 
the statistical distribution of the particle centers, and after applying a subsequent dilation process, 
the original microstructure is recovered with greater fidelity (Figures \ref{figone} and \ref{figonestats}). 

We quantify the accuracy of the reconstruction procedure by measuring some \emph{unconstrained}
correlation function not targeted by the reconstruction algorithm.  In Figure \ref{figonestats}, we compare
the lineal path function $L(z)$, defined as the probability of finding a connected linear segment 
of length $z$ contained completely in the reference phase, for the original donut structure, 
the direct $S_2$-$C_2$ reconstruction, and our reconstruction using erosion and dilation processes.
The direct reconstruction algorithm is not able to capture accurately this type of short-range
continuous connectedness information, accounting for the poorer reconstruction in Figure \ref{figone}. 
However, the lineal path functions of the original system and the
reconstructed material utilizing erosion and dilation processes match closely, implying
that this information is already \emph{implicitly} encoded in the 
spatial distribution of the donut centers.  By utilizing effective two-point correlation functions 
that directly probe the cores of the donut inclusions, we are able to reduce the topological 
complexity associated with the reconstruction procedure without losing fundamental information
about the connectedness properties of the microstructure.  

\begin{figure}[th]
\centering
$\begin{array}{c c c}
\underset{\text{\Large (a)}}{\includegraphics[width=0.3\textwidth]{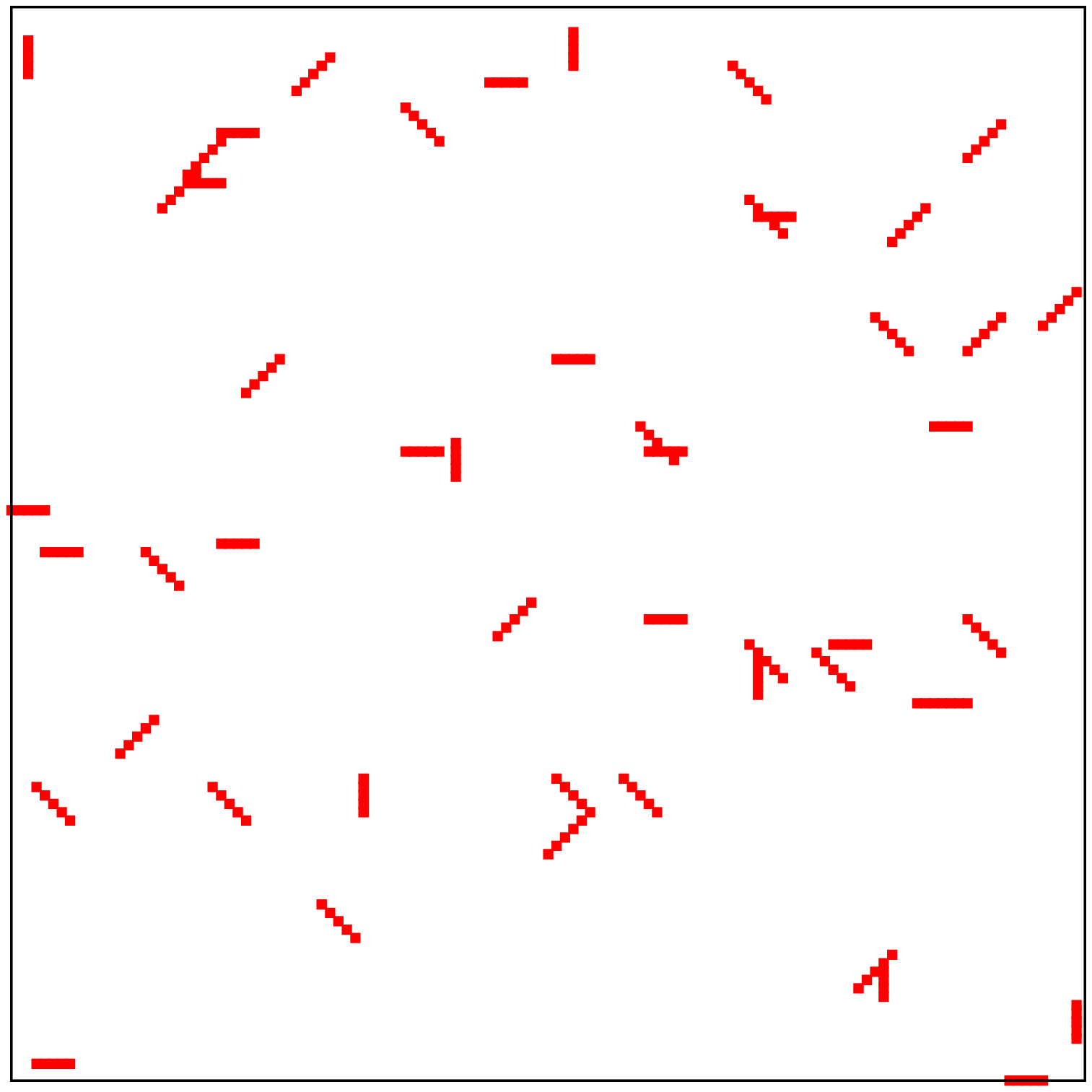}} &
\underset{\text{\Large (b)}}{\includegraphics[width=0.3\textwidth]{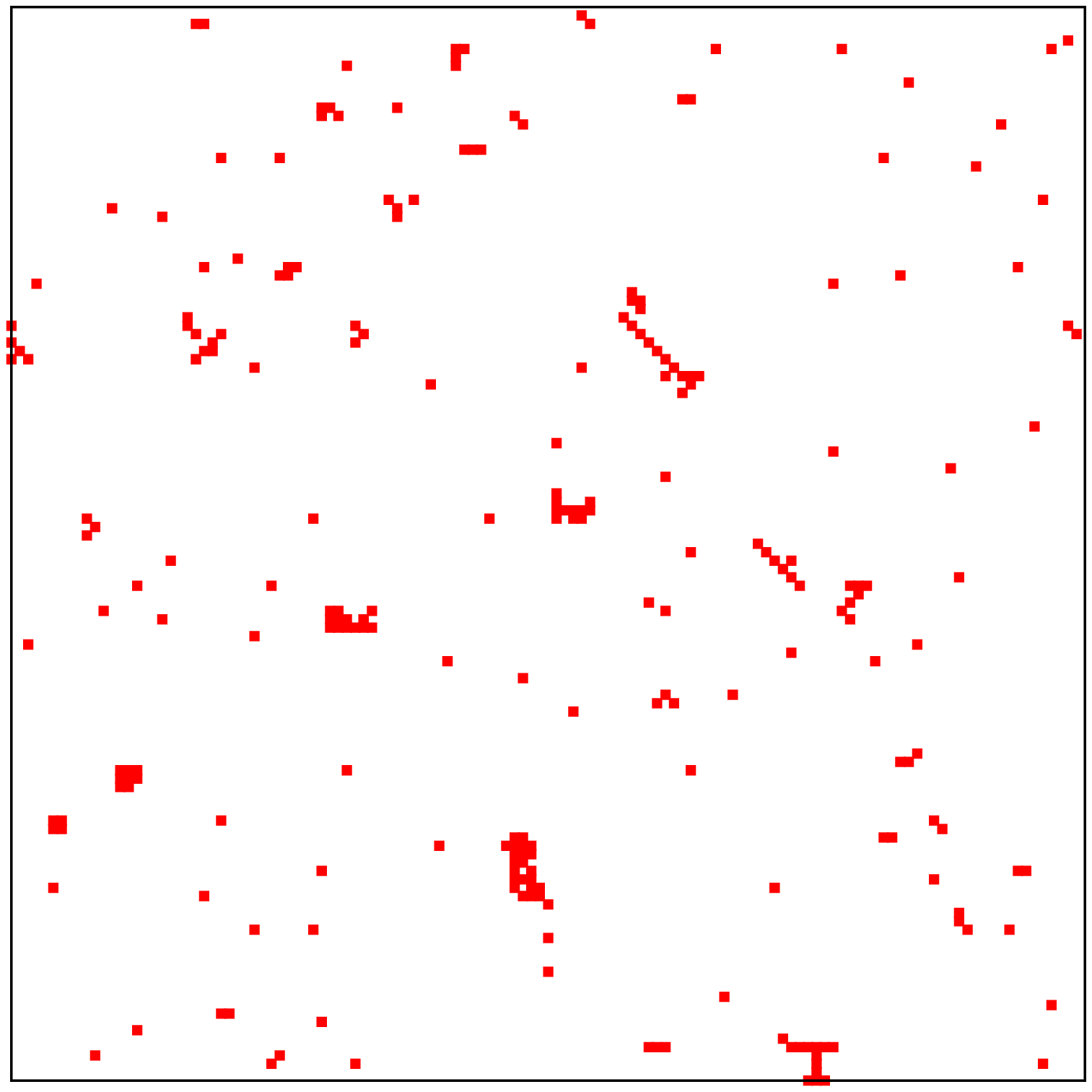}} &
\underset{\text{\Large (c)}}{\includegraphics[width=0.3\textwidth]{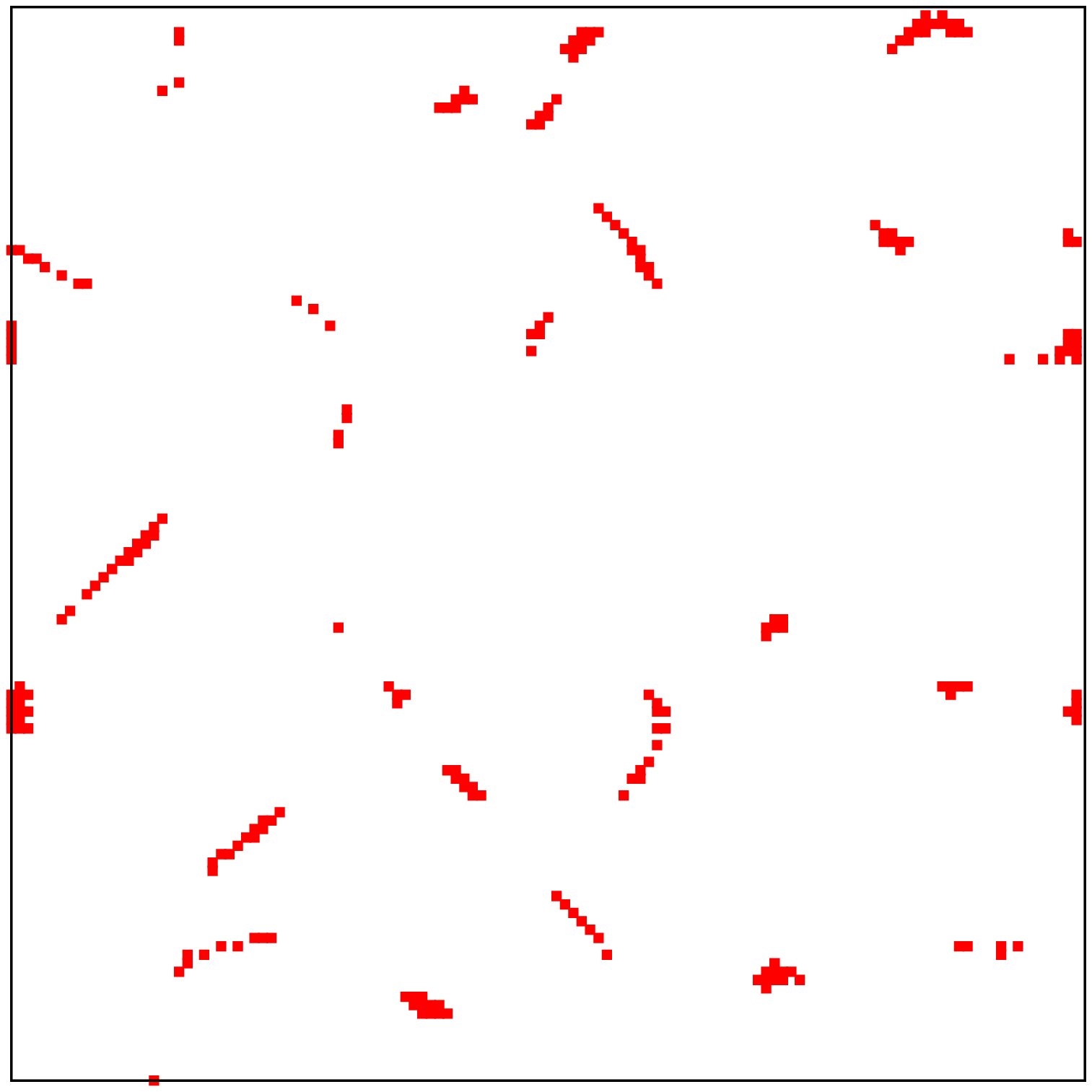}}
\end{array}$
\caption{(Color online)  (a)  Target microstructure consisting of a dilute distribution of cracks
of nearly-zero measure.  (b)  Direct $S_2(r)$-$C_2(r)$ reconstruction of the 
target cracked structure.  (c)  $S_2(r)$-$C_2(r)$ reconstruction 
of the dilated target microstructure followed by an erosion process.  All reconstructions
have a final energy (error) at most $E \sim 10^{-8}$.}\label{filaments}
\end{figure}
We have also applied our methodology to the reconstruction of a dilute distribution of ``cracks'' 
of nearly-zero measure, 
shown in Figure \ref{filaments}.  Note that this prototypical cracked material
is very difficult to reconstruct directly, even utilizing the two-point cluster function (see 
Figure \ref{filaments}b).  Although the
cracks are given a finite size by the pixelation procedure, the volume fraction associated with 
the cracks is small, and a direct $S_2$-$C_2$ reconstruction preserves neither the cracked 
elements of the target structure nor the void-space distribution.  

\begin{figure}[th]
\centering
$\begin{array}{c c}
\underset{\text{\Large (a)}}{\includegraphics[height=2.5in]{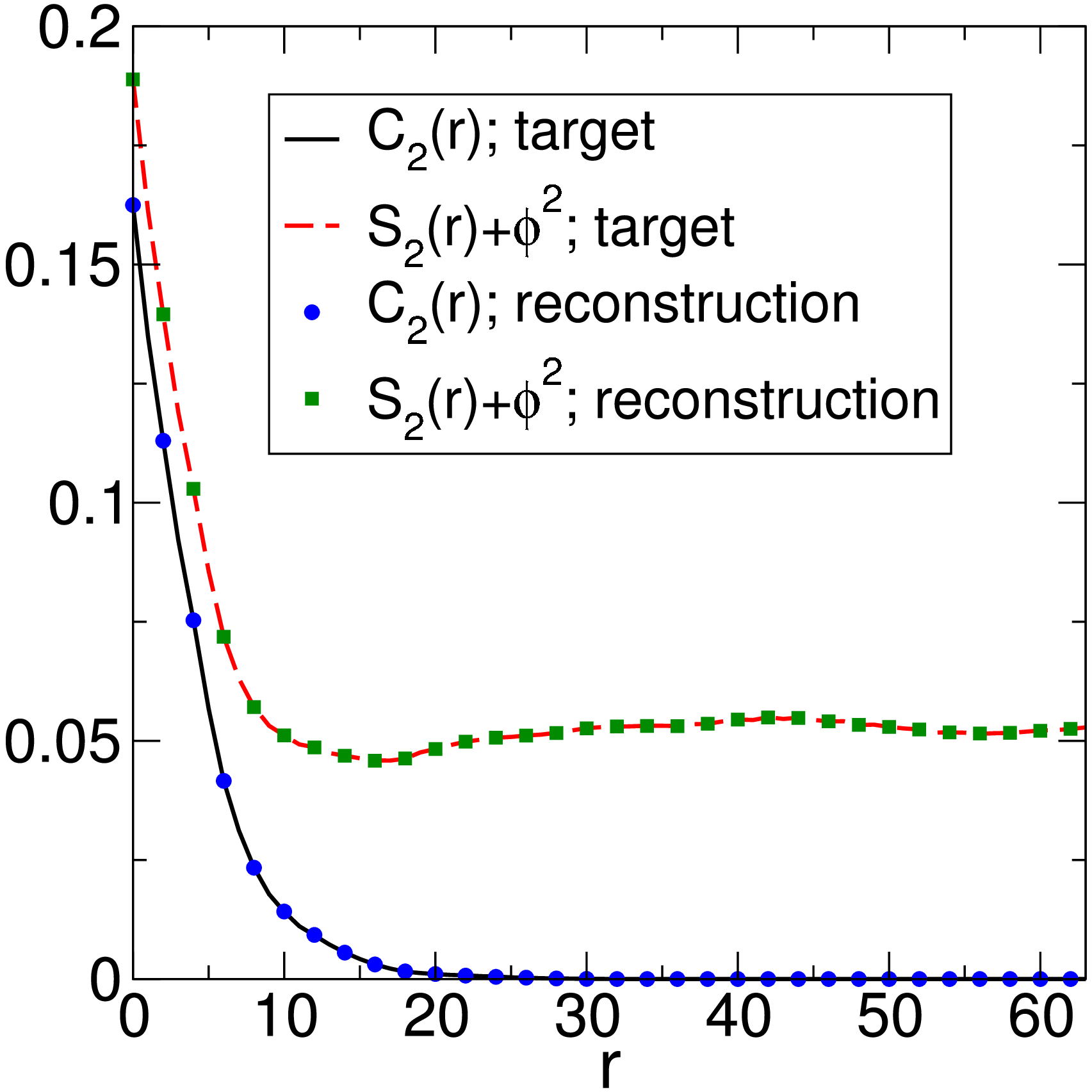}} &
\underset{\text{\Large (b)}}{\includegraphics[height=2.5in]{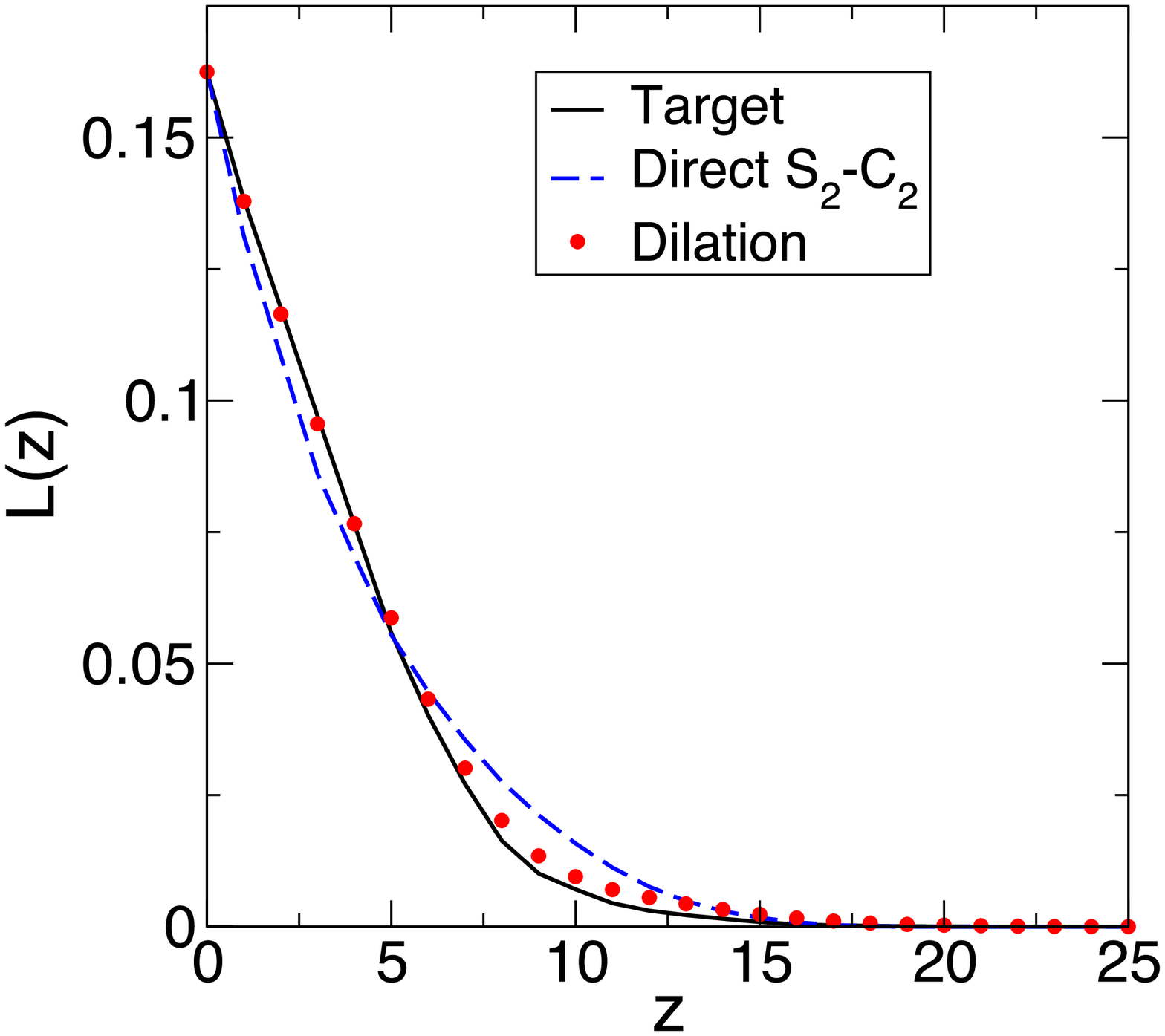}}
\end{array}$
\caption{(Color online)  (a)  $S_2(r)$ and $C_2(r)$ for the dilated target cracked structure and its reconstruction.
(b)  Lineal path functions for the dilated target cracked structure, its reconstruction, 
and the dilation of the direct $S_2(r)$-$C_2(r)$ reconstruction of the initial target 
microstructure.}\label{filS2C2}
\end{figure}
However, by dilating the target structure
and reconstructing the dilated system, we have been able to improve the accuracy of the reconstruction 
procedure.  Figure \ref{filS2C2}  shows excellent agreement between the two-point probability
and cluster functions for the dilated target and reconstructed microstructures, and calculation of the 
lineal path functions demonstrates that the dilation methodology provides quantitatively 
substantial improvement in the accuracy of the reconstruction.  As previously mentioned, 
extended filamentary structures arise in studies of the large-scale structure of the 
Universe \cite{Pe93, Sp05}, and our results therefore suggest that two-point pore-size correlation functions
contain important structural information about such systems, beyond even the standard
two-point probability function $S_2$ and the two-point cluster function $C_2$.

\section{Discussion and Conclusions}

We have introduced a different reconstruction procedure that utilizes two-point pore-size 
correlation functions embodied in $H_2$
 in conjunction with clustering information to improve reconstructions
of two-phase random heterogeneous media. 
It is noteworthy that these two-point functions and the canonical
$n$-point function $H_n$ are generalizations of the Minkowski
functionals \cite{Ar10}.
Our results support the important notion that 
the void space is fundamental to the microstructural properties of a heterogeneous system 
\cite{To10,Za11}.   Our methodology provides a natural extension of the 
Jiao-Stillinger-Torquato reconstruction algorithm and the correlation functions used therein \cite{Ji09}.
A major conclusion of this earlier study was that the two-point cluster function $C_2$ 
employed actually embodies higher-order structural information in a way that makes it a highly
sensitive statistical descriptor over and above the standard two-point function $S_2$.
Thus, $C_2$ has the ability to
“leapfrog” past the usual approach of incorporating additional
information via higher-order versions of $S_2$ (i.e, $S_3$, $S_4$, etc.)
\cite{To83}.
The present work clearly distinguishes the two-point probability and cluster functions $F_2$
and $K_2$, respectively, as highly sensitive descriptors of random media, since they are
generalizations of $S_2$ and $C_2$ and hence
contain even more information than the combination of $S_2$ and $C_2$.

The high information content of $F_2$ and $K_2$ suggests that it would be profitable
to explore their use to characterize the structure and physical properties of not only random media,
but molecular systems. For example, these functions might provide sensitive structural signatures of glassy states of matter (an issue of current interest \cite{Lub06})
and improved rigorous upper and lower bounds on the effective properties of a heterogeneous material
\cite{To02}.   These and related issues are
beyond the scope of the present work, and hence represent
interesting future research topics.

As noted earlier,
two-phase media with percolating phases are difficult to target directly with $C_2$, which becomes long-ranged
at the percolation threshold.  However, preliminary reconstruction results for such cases obtained by applying dilation and erosion 
processes to generate a nonpercolating target medium, which is then reconstructed 
using the generalized $S_2$ and $C_2$, are promising and will be discussed further in future work.

Although we have focused in this paper on reconstructions of two-dimensional
heterogeneous media in order to justify our methods, previous work on 
reconstructions using $S_2$ and $C_2$ by 
Jiao, Stillinger, and Torquato \cite{Ji09}  suggests that our methods can be easily
extended to higher dimensions.  One interesting three-dimensional application of our work
is the reconstruction of biconnected media, in which both phases percolate simultaneously.
This behavior, though difficult to realize in two dimensions, is more common in three dimensions \cite{Be87,To02},
with important implications for the effective properties of a heterogeneous material \cite{To02}.


\begin{acknowledgments}
This work was supported by the Office of Basic Energy Sciences,
Divsions of Materials Sciences and Engineering, under Award DE-FG02-04-ER46108.
\end{acknowledgments}

\end{document}